\documentclass[epj]{svjour}
\usepackage{graphicx}
\usepackage{textgreek}
\usepackage{hyperref}
\hypersetup{colorlinks = true, allcolors = blue}
\usepackage[utf8]{inputenc}
\usepackage{authblk}
\usepackage{hepparticles}
\usepackage{hepunits}
\usepackage{hepnames}

\begin{document}
\title{The \texttau\ challenges at FCC-ee}
\author{Mogens Dam
}                     
\offprints{}          
\institute{Niels Bohr Institute, Copenhagen University, Denmark}
\date{Received: \today / Revised version: \today }
%
\abstract{
At FCC-ee, about $1.7 \times 10^{11}$ \mbox{Z $\to$ \texttau$^+$\texttau$^-$} events will be produced. This high statistics in the clean e$^+$e$^-$ environment opens the possibility of much improved determinations of \texttau-lepton properties and, via the measurement of the \texttau\ polarisation, of the neutral-current couplings of  electrons and $\tau$s. Improved measurements of \texttau-lepton properties -- lifetime, leptonic branching fractions, and mass -- allow important tests of lepton universality. The experimental challenge is to match as far as possible statistical uncertainties generally at the $10^{-5}$ level. This applies in particular to the lifetime measurement, which is derived from the \mbox{2.2-mm} \texttau\ average flight distance, and for the branching fraction and polarisation measurements, where the cross-channel contamination is of particular concern. These issues raise strict detector requirements, in particular, on the accuracy of the construction and alignment of the vertex detector and of the precise calorimetric separation and measurement of photons and \textpi$^0$s in the collimated  \texttau\ decay topologies.
\PACS{
      {PACS-key}{describing text of that key}   \and
      {PACS-key}{describing text of that key}
     } 
} 
\maketitle

\section{Introduction}

At an instantaneous luminosity of $2.3 \times 10^{36}$\,cm$^{-2}$s$^{-1}$ at two (possibly four) interaction points, the FCC-ee will produce about $5\times 10^{12}$ Z decays during a four-year high-statistics scan around the Z pole \cite{FCC:2018evy,Abada:2019lih,Blondel:2021ema}.
This will result in about $1.7 \times 10^{11}$ decays \mbox{Z $\to$ \texttau$^+$\texttau$^-$}, and opens the door to a very rich \texttau-physics programme, as recently pointed out \mbox{in \cite{Pich:2020qna}}. 
Included are unique, high-precision measurements of the Z-to-\texttau\ couplings 
as well as measurements of a wide palette of \texttau-lepton properties in general. With the experimental advantage of a sizeable, well-defined \texttau-lepton boost-factor, $\beta\gamma \simeq 26$, 
very clean background conditions, and somewhat larger statistics than at \mbox{Belle II}~\cite{Belle-II:2018jsg} and at a potential future high-luminosity \texttau-charm factory~\cite{BNPI-STCF,Luo:2018njj}, complementary and very competitive measurements can be expected.

The \texttau\ lepton  is a compelling probe in precision tests of the Standard Model (SM)
because of the
well-understood mechanisms that govern its production and decay. In the SM, 
the electroweak couplings 
of the three generations of
leptons are universal, and 
the three lepton family numbers are individually conserved. This conservation 
is violated in the neutral sector by the observation of neutrino
oscillations. Via loop diagrams, this induces also lepton flavour violation
among charged leptons. The rates of such processes are however 
unobservably small at the $10^{-50}$ level~\cite{Illana:1999ww}, so that any observation of charged lepton flavour violation,
in the same way as any observation of the violation of lepton universality,
would be an unambiguous signal for physics beyond the SM.

The FCC-ee \texttau-physics programme includes the precise determination of the neutral-current chiral couplings of the electron and \texttau{} via measurement of the \texttau\ polarisation; a much improved test of 
the universality between lepton generations of the charged-current couplings  
via measurement of \texttau-lepton properties, lifetime, mass, and leptonic branching fractions; and very sensitive tests of charged lepton flavour violations in \texttau\ and in Z decays.
Experimental challenges are numerous in order to exploit the rich data sample to the maximum. 
Careful design of the detector systems is necessary to enable control of systematic effects down as far as possible towards the targets defined by the
statistical precisions.
The collimated \texttau-decay topologies call, in particular, for 
a very good multi-particle separation both for charged and for neutral final-states particles. In general, high performance vertexing, tracking and electromagnetic calorimetry will be essential. 

With focus on the measurement methodology and on the associated detector requirements, three main topics are considered in this essay: \emph{i})~measurement of the \texttau\ polarisation and of exclusive \texttau\ branching fractions, \emph{ii})~test of lepton universality
via precision measurements of \texttau-lepton properties, and \emph{iii})~tests
of charged lepton flavour violations in Z and in \texttau\ decays. 
A wider discussion of detector requirements at FCC-ee can be found in Ref.\ \cite{Azzi:2021ylt}.


\section{\texttau\ polarisation and \texttau\ exclusive branching fractions}

The \texttau\ polarization in Z decays is one of the most sensitive electroweak observables~\cite{ALEPH:2005ab}. The analysis of the polarization's dependence on the scattering 
angle $\theta$ between the \texttau$^-$ and the e$^-$ beam
gives access to both the \texttau\ and electron chiral
coupling asymmetries $\mathcal{A}_\text{\texttau}$ and $\mathcal{A}_\mathrm{e}$ independently,
\begin{equation}
  P(\cos\theta) =
  \frac{\mathcal{A}_\text{\texttau}(1+\cos^2\theta)+2\mathcal{A}_\mathrm{e}\cos\theta}
       {(1+\cos^2\theta)+2\mathcal{A}_\mathrm{e}\mathcal{A}_\text{\texttau}\cos\theta},
\end{equation}
and serves as a crucial ingredient  of a full lepton-by-lepton extraction of
the neutral-current chiral couplings. In the leptonic decays,
$\text{\texttau} \to \text{e} \bar{\text{\textnu}} \text{\textnu}$,
$\text{\textmu} \bar{\text{\textnu}} \text{\textnu}$,
and in decays to a single final-state hadron,
$\text{\texttau} \rightarrow h \text{\textnu}$, 
$h=\text{\textpi, K}$, the polarization is derived from the charged-particle momentum distribution.
For decays with multiple final-state hadrons, more variables are exploited. As
an important example, for the dominant decay mode,
$\text{\texttau} \to 
\text{\textpi\textpi}^0\text{\textnu}$, the two
sensitive variables are the \textpi\textpi$^0$-system total energy and the asymmetry
between the \textpi\ and \textpi$^0$ energies.
With each channel having its own set of variables with different dependence
on the polarization, a clean separation between channels, e.g.\ between
$\text{\texttau} \to \text{\textpi\textnu}$ and
$\text{\texttau} \to \text{\textpi\textpi}^0\text{\textnu}$,
is essential. 

It is important to realise that not only does one have to 
separate channels according to \textpi$^0$ multiplicity, but one also has to identify and measure precisely the individual \textpi$^0$s.
%
The collimation of final states 
tends to complicate \textpi$^0$
reconstruction, since photons are close to one another and/or close to showers generated by charged hadrons. The two photons from a high-energy \textpi$^0$ may merge due to the limited spatial resolving power of the ECAL. On the other hand, fake photons may occur due to hadronic interactions or fluctuations of electromagnetic showers.
Detailed simulation studies are called for in order to quantify precisely the requirements on the ECAL. Different calorimeter designs have been proposed each with different performance charateristics in terms of transverse and longitudinal granularity and
energy and spatial resolutions~\cite{R1calo}. Studies should allow a comparison and optimisation of these designs.

An experimentally related measurement is that of the \texttau-decay branching
fractions. Experience from LEP~\cite{ALEPH_TauBf05} shows that the branching
fractions are most precisely determined via a global analysis method, where, in
a preselected \texttau$^+$\texttau$^-$ event sample, all decays are categorised
concurrently according to a set of predefined decay modes. An important outcome
of the analysis is a precise determination of the vector and axial-vector
spectral functions, which provide important information for the
extraction of $\alpha_\mathrm{s} (m^2_\text{\texttau})$ and $\alpha_\mathrm{QED} (q^2)$.

For a complete determination of the \texttau-decay branching fractions,
K/\textpi\ separation~\cite{Wilkinson:2021ehf} is needed over the full allowed momentum range of the final state mesons, i.e.\ from the lowest observable momentum to the beam momentum. Whether this is also the case for the precise
\texttau-polarization measurement has to be understood. 
Due to the different masses of the \textpi{} and K mesons, the kinematics of the \textpi\textnu\ and K\textnu{} (\textrho\textnu{} and  K$^*$\textnu) final states obviously differ. And since the polarisation is extracted from the kinematic distributions, it will be necessary to know the relative proportions of \textpi\textnu\ relative to K\textnu{} (\textrho\textnu{} relative to K$^*$\textnu) in the selected samples. For experiments with no K/\textpi\ separation capabilities, this would have to rely on branching-fraction measurements from other experiments and/or facilities.

With the current limit on the still unobserved seven-prong \texttau-decay mode, up to $\mathcal{O}(10^5)$ such decays could be produced at FCC-ee. Kinematically, up to 11 charged pions (plus one \textpi$^0$) are allowed. For the design of the
tracking detectors~\cite{R1tracking}, it is a worthy exercise to make sure that such strongly
collimated, high-multipicity topologies can be correctly reconstructed. The precise,
high-statistics measurement of multiprong \texttau\ decays should bring a much improved determination of the \texttau-neutrino mass, where the current 18.2-MeV limit stems from LEP~\cite{ALEPHnuTauMass}. Also for this measurement, K/\textpi{} separation will be an important asset in selecting heavy final states with one or multiple charged kaons. As an interesting, and experimentally probably very challenging, example, the 3K+2\textpi{} final state has a minimum mass only 17 MeV below the \texttau\ mass.

\section{Tau lepton properties and lepton universality}

Precision measurements of \texttau-lepton properties -- mass, lifetime, and leptonic branching fractions -- provide compelling tests of lepton universality, the expectation that the couplings between the charged lepton and its associated neutrino are equal for the three generations, $g_\mathrm{e} = g_\text{\textmu} = g_\text{\texttau}$.
Firstly, the ratio $g_\text{\textmu}/g_\mathrm{e}$ 
can be readily derived
from the ratio of the branching fractions for the two decay modes,
$\text{\texttau}\to\text{\textmu}\bar{\text{\textnu}}\text{\textnu}$ and $\text{\texttau}\to\mathrm{e}\bar{\text{\textnu}}\text{\textnu}$.
Secondly, the 
ratio, $g_\text{\texttau}/g_\ell$, between \texttau\ and light lepton $\ell=\mathrm{e}, \text{\textmu}$,
can be derived from the relation
\begin{equation}
  \left( \frac{g_\text{\texttau}}{g_\ell} \right)^2 =
  \mathcal{B}(\text{\texttau}\to\ell\bar{\text{\textnu}}\text{\textnu})
  \cdot
  \frac{\tau_\text{\textmu}}{\tau_\text{\texttau}} \cdot
  \left(\frac{m_\text{\textmu}}{m_\text{\texttau}}\right)^5,
\end{equation}
where (small and known) effects due to phase space and radiative and electroweak corrections
have been omitted.
In both cases, current data support lepton universality at the $\mathcal{O}(10^{-3})$ level.

Historically, lepton universality tests took a giant leap with the
LEP measurements based on around one million
$\text{Z} \rightarrow \text{\texttau}^+\text{\texttau}^-$ events (for a complete set of references, see \cite{ParticleDataGroup:2020ssz,HFLAV:2019otj}).
Whereas precise \texttau-mass measurements (currently
$\mathcal{O}(10^{-4})$ precision) have been dominated by threshold scans, LEP
provided important measurements of both the 
lifetime and the leptonic branching fractions. Indeed, the LEP branching
fraction measurements ($\mathcal{O}(10^{-3}$)) still stand unchallenged, whereas
the world-average life-time measurement ($\mathcal{O}(10^{-3}$)), since LEP,
has seen an improvement by about a factor two from a high-statistics Belle 
measurement~\cite{Belle_TauLT14}.

At FCC-ee, with the large number of \texttau\ decays available for study, 
statistical precisions on \texttau-lepton properties will be generally at the $10^{-5}$ level. 
Detailed studies are needed in order to identify the key detector requirements that will allow systematic uncertainties to follow down as far as possible towards this target.
At this stage, a list of observations are:
\begin{description}
\item[\textbf{Lifetime:}] At Z-pole energies, the \texttau\ lifetime is determined via
  measurement of the 2.2-mm average flight distance. A lifetime measurement matching
  the 
  statistical precision would then correspond to a flight-distance measurement to a few tens of nanometers accuracy. Approaching towards this target imposes formidable
  requirements on the construction and alignment accuracy of the vertex
  detector. With a 10--15-mm beam pipe radius, the first vertex detector
  layer will be very close to the beam line, and an impact-parameter resolution
  of about 3\,\textmu m is projected. This is a factor five smaller than at LEP, and combined with the possibility of much improved systematics checks from 
  the enormous event samples, a substantial improvement with respect to LEP should be certainly within reach.
\item[\textbf{Leptonic branching fractions:}] At Z-pole energies, where the separation of a clean and relatively unbiased
  sample of $\text{\texttau}^+\text{\texttau}^-$ events has been
  demonstrated,
  accurate measurements of the \texttau\ leptonic branching fractions rely
  primarily on the ability to separate precisely
  the two leptonic decay modes, on the one hand, from the single-prong
  hadronic modes, on the other. The separation relies critically on a
  fine-grained calorimeter system combined with a dedicated muon system~\cite{R1muondet}. In order to reach the required precise understanding of efficiencies and backgrounds, an independent means of particle identification will undoubtedly 
  prove indispensable. In this respect, it is important to realise that the precise LEP results came from experiments were powerful $\mathrm{d}E/\mathrm{d}x$ measurements played an important r\^{o}le for particle identification. For the separation of low-momentum electrons and pions, a ring-imaging Cherenkov detector may also be an import asset.
\item[\textbf{Mass:}] With the very large statistics, an important improvement of the
  \texttau-mass measurement may be possible via the so-called pseudomass
  method pioneered by ARGUS~\cite{ARGUS_TauMass92} and later exploited by OPAL~\cite{OPAL_TauMass00},
  \textsc{BaBar}~\cite{BaBar_TauMass09}, and Belle~\cite{Belle_TauMass07}. In three-prong\ \texttau\ decays, the
  pseudomass variable depends on the measured mass and momentum of the 3\textpi\
  system and on the beam energy. At FCC-ee, the beam energy is controlled
  to a negligible $10^{-6}$ level via resonant spin depolarisation~\cite{Blondel:2019jmp}, and only the
  measurement of the 3\textpi\ system contributes to the uncertainty. As a reference process, in order to control the mass and momentum
  scale, it is suggested to exploit the very large  J/$\psi$ sample
  from Z decays
  ($\mathcal{B}(\text{Z}\rightarrow \text{J}/\psi X) = 3.5 \times 10^{-3}$),
  and the fact that the
  $\text{J}/\psi$ mass is known to the 10$^{-6}$ precision level from a measurement by KEDR~\cite{Anashin:2015rca} at the VEPP-4M collider, likewise based on resonant spin depolarisation.
  It can be also
  considered to make use of \texttau\ decays with higher charged-particle
  multiplicities that will provide a larger fraction of events close to the
  end-point of the pseudomass distribution. 
\end{description}

\section{Charged Lepton Flavour Violation}

With the huge FCC-ee statistics, very sensitive tests of charged lepton flavour violating (cLFV) processes can be performed in Z decays as well as in \texttau\ decays.

\subsection{Z decays}

Searches for flavour violating Z decays into {\textmu}e, {\texttau\textmu}, and
{\texttau}e final states have been performed at 
LEP with the most precise limits from DELPHI~\cite{DELPHI97_CLFVZ}  and OPAL~\cite{OPAL95_CLFVZ}, and, more recently, at
LHC~ by ATLAS~\cite{Aad:2014bca,ATLAS:2021bdj}. With the recent ATLAS updates on the \texttau\ modes~\cite{ATLAS:2021bdj}, all
LEP bounds have now, finally after 25 years, been superseded. Bounds are slightly below $10^{-6}$ for the \textmu{}e mode and slightly below $10^{-5}$ for the two \texttau{} modes.

In $\text{e}^+\text{e}^-$ collisions, the $\text{Z} \to \text{\textmu e}$
process would have the astonishing clean signature of a beam-energy electron in one hemisphere
recoiling against a beam-energy muon in the other. The dominant experimental
challenge is believed to be that of so-called
\emph{catastrophic bremsstrahlung}, a rare process by which a muon radiates
off the major fraction of its energy in the ECAL material. This way, a $\text{Z} \to \text{\textmu\textmu}$ event could fake
the {\textmu}e signature. Control of this effect would to first order rely on
the ECAL energy resolution and on the aggressive
veto on a possible soft muon track penetrating beyond the ECAL.
Moreover, longitudinal ECAL segmentation would allow vetoing of
electromagnetic showers starting uncharacteristically late.
Finally, an independent method of electron/muon separation such as that 
provided by a precise $\text{d}E/\text{d}x$ measurement could potentially be
used to control this effect possibly to a negligible level. Early estimates indicate that an improvement of the current limit by 2--3 orders of magnitude should be within reach.

The pursuit of
$\text{Z} \to \text{\texttau\textmu}\ (\text{{\texttau}e})$ decays amounts
to a search for events with
a \emph{clear tau decay} in one hemisphere recoiling against
a \emph{beam-momentum muon (electron)} in the other. This signature
receives backgrounds from ordinary \texttau\texttau\ final states, where one \texttau\ decays leptonically,
$\text{\texttau} \to \text{\textmu}\bar{\text{\textnu}}\text{\textnu}$
($\text{\texttau} \to \text{e}\bar{\text{\textnu}}\text{\textnu}$),
with the two neutrinos being very soft, and the charged lepton consequently appearing close to the end-point at the beam momentum.
The separation of signal and background therefore depends
on the experimental precision by which a \emph{beam-momentum particle} can be
defined, and hence on the momentum resolution at 45.6 GeV.
Studies have demonstrated~\cite{Dam:2018rfz} that the search sensitivity scales linear in the momentum resolution combined (in quadrature) with the contribution from the spread on the collision energy 
($\delta E_\mathrm{CM}/E_\mathrm{CM} = 0.9 \times 10^{-3}$) arising from the 60 MeV beam-energy spread~\cite{FCC:2018evy}. 
For a realistic momentum resolution of $1.5 \times 10^{-3}$, limits of $\mathcal{O}(10^{-9})$ on the Z branching fractions look to be within reach.

\subsection{\texttau\ decays}

Very stringent tests of cLFV have been
performed in muon decay experiments where branching-fraction limits below
$10^{-12}$ on both of the neutrinoless decay modes
$\text{\textmu}^-\to\text{e}^-\text{\textgamma}$~\cite{MEG_MuToEGamma16}
and $\text{\textmu}^+\to\mathrm{e^+e^+e^-}$~\cite{SINDRUM_MuTo3E88}
have been established. All models
predicting cLFV in the muon sector imply a violation also in the
\texttau\ sector, whose strength is often enhanced by several orders of
magnitude, usually by some power in the tau-to-muon mass ratio. Studying cLFV
processes in \texttau\ decays offers several advantages compared to muon
decays. Since the \texttau\ is heavy, more cLFV processes are kinematically
allowed, and in addition to the modes $\text{\texttau}\to\text{{\textmu}(e)}+\text{\textgamma}$ and
$\text{\texttau}\to\text{{\textmu}(e)}+\ell^+\ell^-$, also semileptonic and fully hadronic modes can be probed. 

A first FCC-ee simulation study~\cite{Dam:2018rfz} was carried out of 
$\text{\texttau}\to\text{\textmu\textgamma}$
and
$\text{\texttau}\to 3\text{\textmu}$ 
as
benchmark modes. The analysis strategy employed a \emph{tag side} to identify a
clear Standard-Model \texttau\ decay and a \emph{signal side} where cLFV
decays were searched for. Search variables employed were the total energy and the invariant mass of the final-state system.
The present $\mathcal{O}(10^{-8})$ bounds on both
modes are set at the $b$ factories~\cite{BABAR_TauToLeptGam10,Belle:2021ysv,BELLE_TauTo3Lept10}.
The study suggests possible improvements of about one (two) orders of magnitude
for the 
$\text{\texttau}\to\text{\textmu\textgamma}$
($\text{\texttau}\to 3\text{\textmu}$)
mode, largely compatible with
similar projections from \mbox{Belle II}~\cite{Belle-II:2018jsg}. 
In particular, the
$\text{\texttau}\to\text{\textmu\textgamma}$ search is background limited,
and the sensitivity depends strongly on the assumed detector resolutions.
Most importantly, the sensitivity was found to scale 
linearly (or slightly stronger) in the photon energy resolution, which was taken as $16.5\%/\sqrt{E\,(\text{GeV})}$, inspired by the typical performance of CALICE-like silicon-based calorimeters~\cite{R1calo}. The photon position resolution is also important, as it contributes to the invariant-mass calculation. 
Here, where a position resolution of 2 mm was assumed, the two contributions from the energy and position resolutions were found to balance equally in the invariant mass calculation.
%
Interesting for this measurement is the recently proposed fine-grained (two longitudinal layers, $1\times 1$ cm$^2$ lateral segmentation) crystal calorimeter~\cite{Lucchini:2020bac}, which would have an extremely good energy resolution of $3\%/\sqrt{E\,(\text{GeV})}$ combined with a spatial resolution of better than 1~mm.



From the two benchmark modes, there is a long way to the nearly 50 cLFV search modes explored primarily by \textsc{BaBar} and Belle and  
summarised in \cite{HFLAV:2019otj}.
More than one third of these modes involve charged
kaons and would depend on the ability to separate pions and kaons over a wide momentum range.

\section{Unexplored Routes involving Muon Polarisation Measurement}

It is interesting to notice that the enormous FCC-ee statistics allows the observation and measurement of a sizeable sample of muons decaying inside the detector volume. Indeed, 1.2 million muons from \mbox{Z $\to$ \textmu\textmu} events will decay, \mbox{\textmu\ $\to$ e\textnu\textnu}, for each meter of flight distance. This opens the possibility of measuring the polarisation of muons from Z decays in the same way as the polarisation of \texttau{}s is measured in the electronic decay mode, \mbox{\texttau\ $\to$ e\textnu\textnu}. 
From the distribution of the electron momentum fraction, $x=p_\text{e}/p_\text{\textmu}$, where the electron energy is precisely measured by the ECAL, a sample of one million decays would result in a statistical precision of $4.5 \times 10^{-3}$. This can be improved by about 20\% by inclusion of the acollinearity angle between the decay electron and the opposite hemisphere muon.

Parallel to this, but experimentally more challenging, is a unique possibility of contributing to the determination of the Lorentz structure of the \texttau-decay amplitude. 
A few hundred thousand muons from \texttau\ decays, \mbox{\texttau\ $\to$ \textmu\textnu\textnu}, will decay, \mbox{\textmu\ $\to$ e\textnu\textnu},
per meter flight distance inside the detector volume. The challenge now is \emph{i}) to recognise in each case that a decay has happened by identifying the decay vertex via a change in the track curvature (and, if available, possibly also in the specific ionisation), and \emph{ii}) to measure the momentum both before (muon momeutum, $p_\text{\textmu}$) and after (electron momeutum, $p_\mathrm{e}$) the decay vertex.
In cases where the decay happens well inside the tracking volume, this may be possible, and from the distribution of momentum fractions, $x=p_\mathrm{e}/p_\text{\textmu}$, the polarisation of the muon may be derived.

\section{Conclusions and Outlook}
\label{section:conclusion}

With its enormous sample of \mbox{Z $\to$ \texttau$^+$\texttau$^-$} events, it is justified to think of FCC-ee (also!) as a dedicated \texttau{} factory. As an example, touched upon above, it opens the possibility to measure the Fermi decay constant in \texttau\ decays with a \emph{statistical} precision of 10$^{-5}$, 
only about one order of magnitude less precise than what is known today from muon decays. Clearly, it would be a formidable, possibly impossible, task to control systematics down to this same level of precision. 
However, it illustrates well the situation and emphasises how a thorough study unearthing all aspects of the experimental possibilities is deserved. 

From the palette of measurements discussed in this essay, strong requirements on the detector design have been already pointed to. Important issues include:
a very precisely constructed and aligned silicon vertex detector for ultimate lifetime measurements; a light and precise tracking system with superior multi-track separation for precise momentum and mass measurements; a fine-grained electromagnetic calorimeter for precise measurement of photons and \textpi$^0$s in dense topologies; a high-performant muon system for clean pion/muon separation; particle identification abilities over the full momentum range for kaon/pion separation and, very importantly, for an independent, non-destructive electron/pion separation.

%
%
%

\bibliographystyle{myutphys}
\bibliography{references}
\end{document}